\documentclass[notitlepage,prl,twocolumn,nofootinbib]{revtex4-1}
\usepackage{graphicx}

\begin{document}

\title{NEOS data and the origin of the 5 MeV bump in the reactor antineutrino
    spectrum}

\begin{abstract}

We perform a combined analysis of recent NEOS and Daya Bay data on the
reactor antineutrino spectrum. This analysis includes approximately
1.5 million antineutrino events, which is the largest neutrino event
sample analyzed to date. We use a double ratio which cancels flux
model dependence and related uncertainties as well as the effects of
the detector response model. We find at 3--4 standard deviation
significance level, that plutonium-239 and plutonium-241 are
disfavored as the single source for the the so-called 5\,MeV
bump. This analysis method has general applicability and in particular
with higher statistics data sets will be able to shed significant
light on the issue of the bump. With some caveat this also should
allow to improve the sensitivity for sterile neutrino searches in
NEOS.
\end{abstract}

\author{Patrick Huber}
\email{pahuber@vt.edu}
\affiliation{Center of Neutrino Physics, Virginia Tech, Blacksburg, USA}
\date{\today}
\maketitle


The 5\,MeV bump in the reactor antineutrino spectrum was first
reported by the RENO collaboration~\cite{Seo:2014xei,RENO:2015ksa} and
confirmed by Daya Bay~\cite{An:2015nua} and Double
Chooz~\cite{Abe:2015rcp}. Several nuclear physics hypothesis have been
put forward to explain the origin of the
bump~\cite{Dwyer:2014eka,Zakari-Issoufou:2015vvp,Hayes:2015yka,Sonzogni:2016yac,Rasco:2016leq}.
Some argue that the bump could be caused by a particular fissile
isotope, {\it e.g.} uranium-238~\cite{Hayes:2015yka}, whereas other
explanations focus on one specific or a small number of fission
fragments.  Experimentally identifying whether one isotope is
predominantly responsible for the bump or not is a crucial step. The
NEOS collaboration recently presented first results~\cite{ICHEP} based
on a single-volume, gadolinium-doped, liquid scintillator antineutrino
detector of approximately 1 ton fiducial mass at 24\,m distance from
the core of a power reactor. This measurement constitutes another
piece in this puzzle.

\begin{figure*}
\includegraphics[width=\textwidth]{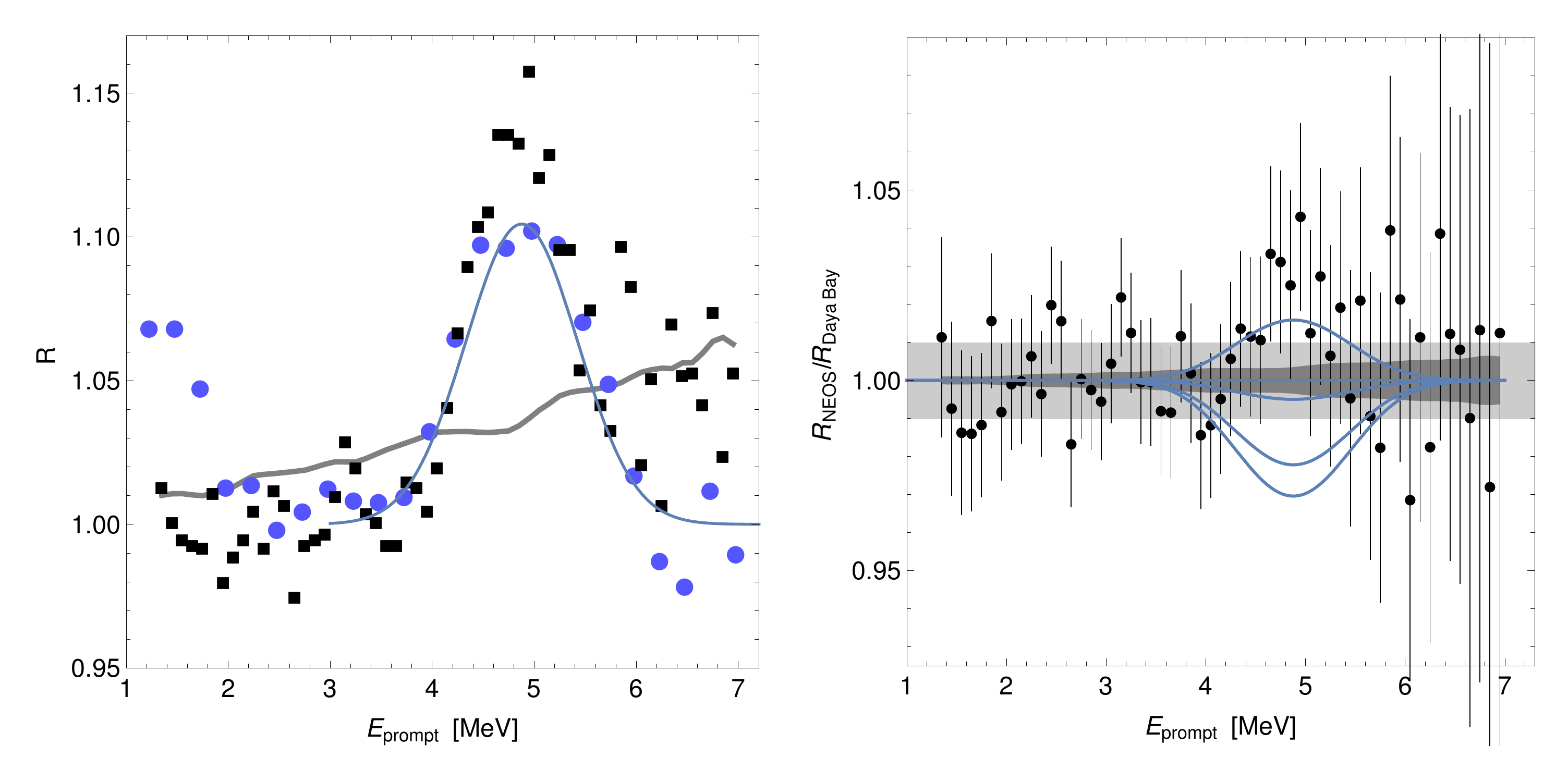}
\caption{\label{fig:double} In the left hand panel, the ratio $R$ of
  measured spectrum to the Huber+Mueller prediction using the
  respective fission fractions is shown for Daya Bay as blue circles
  and for NEOS as black squares. The dark gray line is the correction
  applied to the NEOS data arising from the different fission
  fractions in both experiments according to the Huber+Mueller
  model. The blue curve corresponds to the best fit bump shape in Daya
  Bay. In the right hand panel, the black dots show the double ratio
  $R_\mathrm{NEOS}/R_\mathrm{Daya Bay}$, the error bars are the purely
  statistical errors from NEOS and the nuisance parameters
  $\mathbf{\eta}$, defined in the text, are at their best fit
  values. The various lines depict the predictions for the bump in the
  double ratio. The light gray band corresponds to the uncertainty
  stemming from the NEOS detector response, whereas the dark gray band
  is the uncertainty of the fission fraction correction and corresponds
  to 10\% of the correction, shown as dark gray line in the left hand
  panel. }
\end{figure*}

The basis for the following analysis is the fact the Daya Bay
measurement and the NEOS result were obtained at different effective
fission fractions for uranium-235, uranium-238, plutonium-239 and
plutonium-241, respectively. The fission fractions in Daya Bay are
0.561, 0.076, 0.307 and 0.056~\cite{An:2016srz}, whereas for NEOS they
are 0.655, 0.072, 0.235, 0.038~\cite{OhPrivate}. The bump in the Daya
Bay data is well described in prompt energy by a Gau\ss ian with
central value of 4.9\,MeV, a width of 0.55\,MeV and an amplitude of
10.4\%, as can be seen in the left hand panel of
Fig.~\ref{fig:double}. If the bump were equally caused by all four
fissile isotopes it should have the same amplitude in both data
sets. On the other hand, if the bump is for instance only due to
uranium-235, the bump in NEOS should be $0.655/0.561=1.17$ times
larger, and similarly for all the other fissile isotopes.

The analysis of the bump in either experiment usually relies on a
comparison with the Huber+Mueller flux
model~\cite{Mueller:2011nm,Huber:2011wv}, however this flux model has
large uncertainties in itself which limit the obtainable accuracy
significantly~\cite{Huber:2016fkt}. Here, we will try to directly
compare the Daya Bay spectrum with the NEOS spectrum. Daya Bay has
reported a spectrum result which has been ``cleaned'' of all detector
effects by unfolding~\cite{An:2016srz}, on the other hand NEOS has
presented a result in prompt energy only. The prompt energy in a
detector using inverse beta decay is given by
$E_\mathrm{prompt}=E_{\bar\nu}-0.8\,\mathrm{MeV}$ since an antineutrino
of 1.8\,MeV, that is at threshold, creates a positron at rest, which
will annihilate and deposit twice the electron mass in gamma rays
$\sim 1\,\mathrm{MeV}$ in the detector.

For a detector as small as NEOS, energy containment of the positron
itself and the 511\,keV annihilation gamma-rays is a major issue and
thus the relation between prompt and neutrino energy is complex. We
try to address this issue by using a double ratio: NEOS not only has
published a prompt event spectrum but also the ratio of the measured
prompt spectrum to the prompt spectrum predicted by the Huber+Mueller
model, $R_\mathrm{NEOS}$, shown as black squares in the left hand
panel of Fig.~\ref{fig:double}. For the Daya Bay unfolded spectrum it
is trivial to compute the corresponding ratio, $\tilde R_\mathrm{Daya
  Bay}$ and it is shown as blue circles in the left hand panel of
Fig.~\ref{fig:double}. However this ratio is for a different set of
fission fractions. We correct for the difference in fission fractions
using the Huber+Mueller model. The resulting correction is small, less
than 5\% of the total flux and manifests itself as linear slope
without any features, as shown as the thick gray line in the left-hand
panel of Fig.~\ref{fig:double}. Assigning a generous bin-to-bin
uncertainty of 10\% to the Huber+Mueller prediction, the effect on the
ratio will be $5\%\times10\%\leq 0.5\%$, as depicted by a dark gray
region in the right-hand panel of Fig.~\ref{fig:double}.  Therefore
the large model uncertainties are greatly reduced and can be neglected
in the following. We will call the corrected ratio $R_\mathrm{Daya
  Bay}$.

We now can form a double ratio $R_\mathrm{NEOS}/R_\mathrm{Daya Bay}$,
which in the absence of detector effects would lead to a complete
cancellation of the flux model up to the negligible correction for the
difference in fission fractions. The NEOS detector response function
is unknown to us, but we can make an educated guess towards its
general properties and can show that only a small correction
results. In Ref.~\cite{ICHEP} the NEOS collaboration quotes an energy
resolution of $5\%$ at 1\,MeV and from the information in
Ref.~\cite{ICHEP} we can infer an energy resolution of approximately
this form
\begin{equation}
\label{eq:sigma}
\sigma(E_\mathrm{prompt})=(0.05\sqrt{E_\mathrm{prompt}}+0.12)\,\mathrm{MeV}\,.
\end{equation}

There are two types of energy losses through the surface: escaping
511\,keV gamma rays and the positron itself can escape as well. Each
process has a mean range $\lambda$ and only events which are closer
than $\lambda$ to the surface can experience a significant energy
loss.

The mean range of a 8\,MeV positron, $\lambda_{e^+}$, in scintillator
is approximately 4\,cm. For a spherical detector of
$1\,\mathrm{m}^3$ volume\footnote{This corresponds to roughly 1 metric ton of
  liquid scintillator.}, approximately 15\% of the volume lies within 
$\lambda_{e^+}$ from the surface. Assuming that one half of all events
within this $\lambda_{e^+}$ thick shell are leaving the detector
and that for those events, the energy deposited in the detector forms
a flat distribution between 0 and the actual energy of the positron,
we obtain a simple, but conservative model of energy losses for
positrons.

For the the two 511\,keV annihilation gammas the mean range
$\lambda_{\gamma}$, which is due to Compton scattering, is about 8\,cm
and approximately 28\% of the total detector volume is within a shell
of this thickness. Using a simple Monte Carlo simulation based on the
Compton scattering cross section, we find that indeed about one half
of the events generated within a shell of thickness $\lambda_\gamma$
will experience energy loss through the surface and we obtain the
actual energy loss distribution. 

\begin{figure}
\includegraphics[width=1\columnwidth]{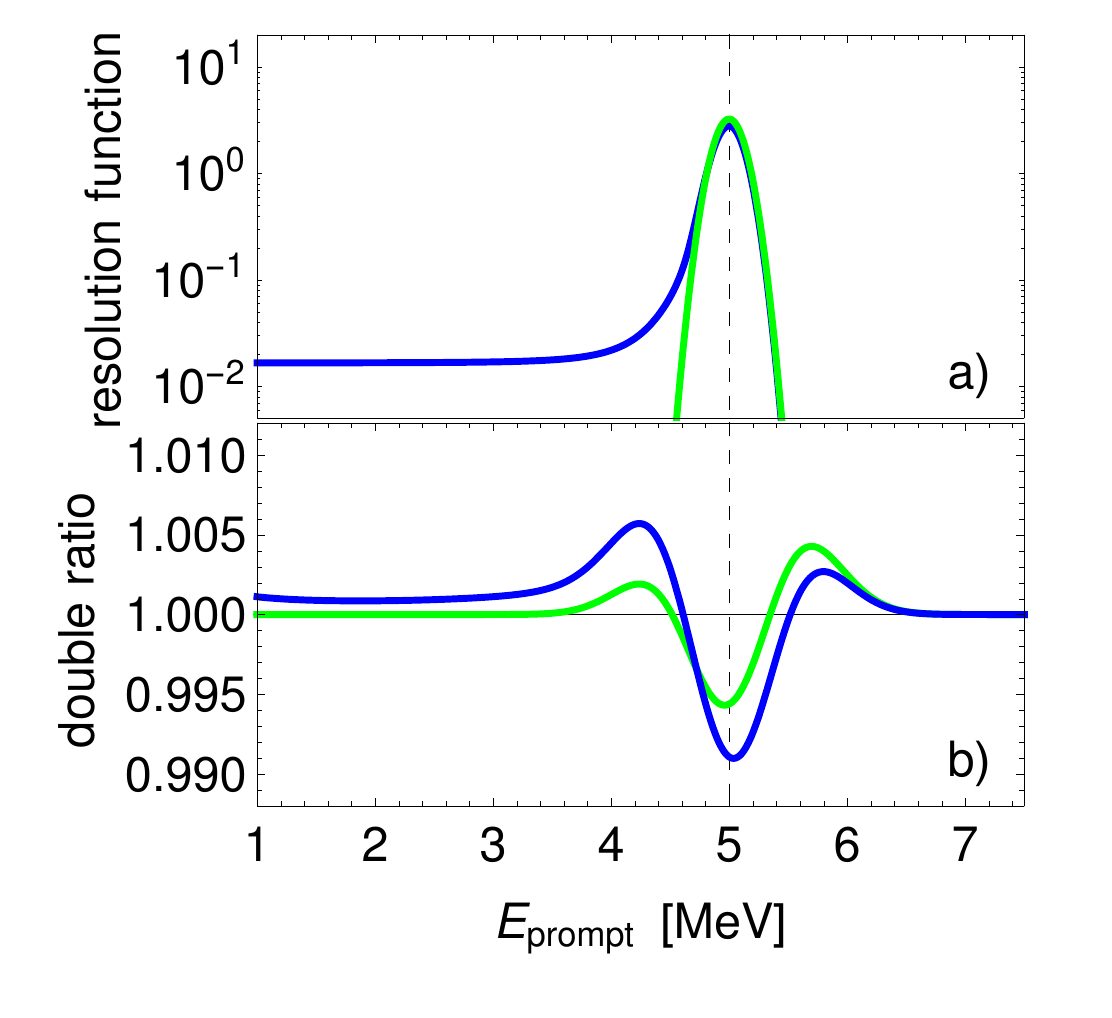}
\caption{\label{fig:cancellation} In panel a) the base-10 logarithm of
  the resolution function for a true prompt energy of 5\,MeV is shown:
  in green (light gray) a simple Gau\ss ian resolution and in blue
  (dark gray) a more realistic detector response as explained in the
  text is used. In panel b) the resulting double ratio is shown for a
  toy example where there is a bump of same amplitude in the data of
  both experiment A and the data of experiment B.  }
\end{figure}

Combining these two ingredients, we obtain our approximate detector
response function $D(E_\mathrm{prompt},E_\mathrm{rec})$, where
$E_\mathrm{rec}$ is the reconstructed energy, which is shown as a blue
line in panel a) of Fig.~\ref{fig:cancellation} and for comparison a
simple Gau\ss ian response function is shown in green. For
$D(E_\mathrm{prompt},E_\mathrm{rec})$ (blue curve) the ratio of events
which have their energy reconstructed below the true energy to those
which have it reconstructed above is 1.32, resulting in a mean
reconstructed energy of 4.78\,MeV for a true energy of 5\,MeV.

We now can study the fidelity of the cancellation of detector effects
with a toy example. Assume experiment A is subject to the detector response $D$
and sees a bump of 10\% amplitude at 5\,MeV. For experiment A we
form the ratio $R_A$ of the experimental mock data (with bump) to a
theory prediction (without a bump) smeared with the same detector
response, $D$. Experiment A corresponds to NEOS. Next, for experiment B we
assume that the data is reported as an unfolded result free from
detector effects and that experiment B does see the same bump. We form
the corresponding ratio $R_B$ of data and theory. In
Fig.~\ref{fig:cancellation} we show the resulting double ratio
$R_A/R_B$ for two different assumptions about the detector response
of experiment A. We see, that we misreconstructed the size of the bump
in terms of the double ratio by less than 0.01. We also see that two
quite different detector response models, a simple Gau\ss ian versus a
model with significant surface energy losses, yield an estimate of the
residual effect within a factor of 2 of each other.  We will include
the 0.01 error in the subsequent analysis despite the fact that this
error is about ten times smaller than the errors resulting from
statistics in NEOS and Daya Bay, thus even if we were to double this
number it would have minor impact on the final result. This
demonstrates the robustness of the double ratio against detector
effects.

Finally, we can look at the resulting double ratio for the actual data
$R_\mathrm{NEOS}/R_\mathrm{Daya Bay}$, which is shown in the form of
black dots in the right-hand panel of Fig.~\ref{fig:double}. Note,
that the absolute flux is left free, so the fact that the double ratio
averages to 1 is of no significance. We clearly observe that the bump
is largely canceled between Daya Bay and NEOS pointing towards the
fact the amplitude and position of the bump in both data sets is very
similar. This double ratio would lend itself extremely well to form
the basis of a sterile neutrino search in NEOS, freeing NEOS largely
from flux model uncertainties, using Daya Bay as far detector for
NEOS. The Daya Bay detectors are at a distance where any sterile
neutrino oscillation, with a $\Delta m^2$ for which NEOS is sensitive,
is averaged out. However, since in a sterile oscillation search many
peaks and valleys can appear as localized features in the data, our
simplistic treatment of the detector response is suspect: the detector
response function nearly cancels for the bump search since the
amplitude \emph{and} position of the bump in both data sets is very
close, see left hand panel of Fig.~\ref{fig:double}. As explained,
this would be not the case for oscillations, since Daya Bay cannot see
any relevant oscillation for the $\Delta m^2$ values in question. The
double ratio for a sterile analysis, therefore presumably requires a
detailed understanding of the detector response and in this case, the
cancellation of flux uncertainties will still apply and be very
beneficial.

In order to quantify the agreement between the two data sets, we
perform a fit to the double ratio of various bump amplitudes. In this
fit we include the statistical errors of the NEOS data and the full
covariance matrix published together with the Daya Bay unfolded
flux~\cite{An:2016srz}. The binning of NEOS data is 100\,keV and thus,
much finer than the binning of the Daya Bay data of 250\,keV. This
makes the direct use of the covariance matrix of Daya Bay somewhat
difficult in a combined fit. We introduce one nuisance parameter
$\eta_i$ for each of the Daya Bay energy bins by multiplying the bin
content with $1+\eta_i$; as these $\mathbf{\eta}$ parameters are
varied in the fit we obtain a shifted Daya Bay spectrum. We then use
linear interpolation on this shifted Daya Bay spectrum to obtain
values for the 100\,keV bins of NEOS. The $\mathbf{\eta}$ parameters
are constrained in the fit by the covariance matrix $\mathbf{V}^{-1}$
by adding the following term to the $\chi^2$-function
$\mathbf{\eta}\mathbf{V}\mathbf{\eta}^T$. The only physical fit
parameter is the bump amplitude in the double ratio where we keep the
bump position fixed at 4.9\,MeV. Note that black data points shown in
Fig.~\ref{fig:double} haven been shifted corresponding to the values
of $\eta_i$ found at the best fit point and the error bars shown are
the statistical errors of the NEOS data set only.

We use this analysis to test the following five hypotheses: the bump
is only in uranium-235 or in uranium-238 or in plutonium-239 or in
plutonium-241, and fifth, the bump is equal in all isotopes. The
result is shown in Tab.~\ref{tab:result}, the best fit is obtained for
$R_\mathrm{NEOS}/R_\mathrm{Daya Bay}$ at the bump of 1.022 with a
$\chi^2$ of 46.7 and the resulting overall goodness of fit is 80\%,
taking 57-1 degrees of freedom. The standard deviation in
Tab.~\ref{tab:result} is obtained by taking the $\chi^2$-value of each
model, subtract the best fit and then convert the resulting $\Delta
\chi^2$ into standard deviations using a $\chi^2$-distribution with 1
degree of freedom, {\it i.e.} the number of standard deviations is
$\sqrt{\Delta \chi^2}$.
\begin{table}[h!]
\begin{tabular}{cccccc}
Isotope&$^{235}$U&$^{238}$U&$^{239}$Pu&$^{241}$Pu&equal\\
\hline
$R_\mathrm{NEOS}/R_\mathrm{Daya Bay}$&1.021& 0.993& 0.971& 0.960& 1.000\\
$\chi^2$&46.9&51.6&60.3&66.0&49.9\\
$\sigma$&0.34&1.93&3.27&3.92&1.55\\
\end{tabular}
\caption{\label{tab:result} $\chi^2$-values for the the bump being
  caused by a single isotope or in equal parts by all isotopes. The
  fit has 57-1 degrees of freedom and the $\chi^2$-minimum is 46.7 and
  occurs at a value of the double ratio of 1.022.}
\end{table}
The case of the bump being caused by either uranium isotope or equally
by all isotopes is clearly preferred over the case where the plutonium
isotopes carry sole responsibility. This result is more conclusive than
one would expect from the numbers given in~\cite{Huber:2016fkt}
because the model uncertainties of the Huber+Mueller model cancel in
the double ratio and thus starts to approach the more ideal case of
negligible flux errors~\cite{Buck:2015clx}.

The fact that the plutonium isotopes are disfavored as the sole origin
of the bump is at odds with the possibility put forward in
Ref.~\cite{Hayes:2015yka}, that epithermal fission of plutonium is
responsible for the bump. This explanation would have the advantage
that it naturally explains why the bump is absent in the integral beta
spectra~\cite{VonFeilitzsch:1982jw,Schreckenbach:1985ep,Hahn:1989zr}
since theses measurements were done in a purely thermal neutron
flux. Our analysis would still allow uranium-238 to be the sole origin
of the bump, but due to the small uranium fission fraction the size of
the bump in uranium-238 would have to be of order 2 (!), which should
naturally leave an imprint in the integral beta spectrum for
uranium-238 fission~\cite{Haag:2013raa}, which is not the case. Thus,
the basic riddle of how to accommodate the bump in the antineutrino
spectrum without leaving a trace in integral beta spectra remains. It
is interesting to note that our analysis slightly prefers uranium-235
as the sole source and that recently it was pointed out that the total
inverse beta decay yield of uranium-235 disagrees most with
predictions~\cite{Giunti:2016elf}. Also, recent results reported by
the RENO collaboration seem to indicate a positive correlation of the
bump amplitude with the uranium-235 fission fraction~\cite{RENO},
which is consistent with the results presented here. In combination
this evidence supports, but does not conclusively establish,
uranium-235 as leading contributor to the bump.

Future measurements at research reactors, which exhibit nearly pure
uranium-235 fission, clearly will help to distinguish the remaining
most likely possibilities: only uranium-235 would predict an amplitude
of the bump of about 0.23--0.26, only uranium-238 predicts an
amplitude of 0 and equal contribution from all isotopes predicts 0.14
for the amplitude. Note, that these reactors run at fission fractions
which are quite different from both NEOS and Daya Bay, thus using the
data of NEOS or Daya Bay as a reference will require applying a larger
correction for fission fractions based on the Huber+Mueller model and
thus a larger fraction of the model uncertainties will apply. Also,
these planned experiments will work at a signal of noise ratio very
much worse than that of NEOS, which may reduce the statistical power
of those experiments. Thus, the impact of these measurements for the
question at hand could potentially be significant, but this impact
depends on the yet unknown actual performance of the detectors.

In summary, we have shown that a double ratio of experimental data and
theory predictions allows to cancel the flux model dependence and
related uncertainties and does not rely on accurate modeling of the
detector response. Based on a combined analysis of NEOS and Daya Bay
data we find with respect to the 5\,MeV bump, that the two plutonium
isotopes are disfavored as sole source of the bump at approximately
3--4 standard deviations. The Daya Bay data set used here corresponds
to about 1,200,000 events and the NEOS data set corresponds to 300,000
events, in combination this is the largest number of neutrino events
analyzed jointly to date. It took only 6 months to accumulate the NEOS
data and thus it is conceivable that this data set could quadruple in
size, and by that, reducing the statistical errors in NEOS by a factor
of 2. This could push the ability to distinguish the sole uranium-235
hypothesis from the case of equal contributions from all fissile
isotopes above the 3\,$\sigma$ level and thus would allow to either
establish or to refute uranium-235 as the single most important
contributor to the 5\,MeV bump.

\acknowledgments

This work was supported by the U.S. Department of Energy under award
\protect{DE-SC0009973}. PH would like to thank C.~Mariani for a
careful reading of the manuscript and D.~Dwyer for useful discussions
and the NEOS collaboration for providing the NEOS data in machine
readable format.

\bibliography{./references.bib}

\end{document}